%% file: main.tex
\documentclass[reprint,superscriptaddress,nofootinbib,onecolumn,aps,pra]{revtex4-2}
\input{preamble}

\input{symbols}

\begin{document}

\title{ Positional memory of skyrmions in magnetic bilayers}
\author{Bruno Barton-Singer}
\affiliation{Institute of Applied and Computational Mathematics, FORTH, Heraklion, Crete, Greece}
\affiliation{Department of Mathematics and Applied Mathematics, University of Crete, 70013 Heraklion, Crete, Greece}
\author{Anusree Navallur}
\affiliation{Department of Mathematics and Applied Mathematics, University of Crete, 70013 Heraklion, Crete, Greece}
\author{Stavros Komineas}
\affiliation{Department of Mathematics and Applied Mathematics, University of Crete, 70013 Heraklion, Crete, Greece}
\affiliation{Institute of Applied and Computational Mathematics, FORTH, Heraklion, Crete, Greece}
\date{\today}

\begin{abstract}
We numerically and analytically study the transient dynamics of magnetic skyrmions in synthetic antiferromagnets under a magnetic field gradient.
We consider skyrmions in a bilayer with antiferromagnetic coupling between the layers.
The skyrmions in the two layers move almost perpendicular to the field gradient and the motion is eventually halted with the two skyrmions at a distance from each other.
We find that the skyrmion displacement is proportional to the field gradient, while the time it takes to reach their final position is almost independent of it.
Below a critical magnetic field gradient strength, the system displays an unusual `remembering' dynamics: when the magnetic field gradient is removed, the skyrmions return to their original positions to a high degree of accuracy.
We explain this observation and other quantitative features using a moduli space dynamics approximation.
We further provide an exact treatment of the dynamics that indicates that deviations from exact memory of the skyrmion position can arise.
\end{abstract}

\maketitle

\section{Introduction}
\label{sec:introduction}

Skyrmions are a subject of interest due to their robustness and their particle-like properties \cite{2017_NRM_FertCros,2018_JAP_EverschorKlaeui}.
Their dynamical behavior is dramatically different in antiferromagnets (AFM) and ferromagnets (FM). In response to various forces, ferromagnetic skyrmions move at an angle to the force with a proportional velocity \cite{MalozemoffSlonczewski}, while antiferromagnetic skyrmions often move in the direction of the force with a proportional acceleration, behaving like Newtonian particles \cite{2016_BarkerTretiakov_PRL,2016_APL_JinLiu}.
A system that has attracted interest in the last years are synthetic antiferromagnets (SAF), which consist of ferromagnetic layers that are antiferromagnetically coupled with the neighbouring layers.
The initial motivation was to get access to antiferromagnetic dynamics, but it is gradually becoming clear that this system has properties not accessible either in ferromagnets or antiferromagnets, and thus should be studied in its own right.

Skyrmions in SAF have been experimentally observed in recent works \cite{2020_NatMatter_LegrandCrosFert,2022_NatComm_JugeSisodiaBoulle,2022_AdvFunMatter_ChenSong}.
The dynamical response of SAF skyrmions has been studied for forces that act the same on both layers, such as electric current interacting through the spin-orbit and spin-transfer torques \cite{2017_NatComm_HrabecRohart,2022_PhysRevB_PanigrahyMallickSampaioRohart,2024_Science_PhamSisodiaBoulle}, or anisotropy gradient \cite{2019_NJP_Ang}. This leads to effective skyrmion inertia and an acceleration in the direction of applied force, like the AFM case. More complicated cases have also been considered where changes in skyrmion radius, helicity or ellipticity make the force due to current different on each layer 
\cite{2022_PhysRevAppl_MsiskaRodriguesLeliaertEverschor-Sitte,l2025_arxiv_LeeVelezOtxoaMochizuki}.
A different type of steady-state dynamics is manifested for SAF skyrmions in response to magnetic field gradient which acts exactly oppositely on each layer \cite{2020_JMMM_Zhou}.
However, the full dynamical response has not been investigated.

In this paper, we study the transient response of skyrmions in SAF bilayers to a magnetic field gradient.
We find that this response exhibits unusual features below a critical strength of the gradient set by the coupling between the skyrmions on the two layers.
The skyrmion in the bilayer undergoes a net motion perpendicular to the field gradient and it eventually finds an equilibrium where the skyrmions in the two layers are at a distance from each other.
When the field is switched off, the skyrmions move in the reverse direction and finally return to their original position with a high precision. Thus, the skyrmions exhibit a form of positional memory, and this behaviour is distinct from that of both FM and AFM skyrmions.
We develop a formalism based on moduli space dynamics that describes this transient behaviour and gives the details of the skyrmion trajectories and dynamics.
Furthermore, an exact treatment of the skyrmion dynamics indicates that the return to original position, despite being an extremely good approximation, is not an exact result.

The paper is organized as follows.
In Sec.~\ref{sec:model}, we introduce the SAF model.
In Sec.~\ref{sec:numericalSimulation}, we present numerical simulations.
In Sec.~\ref{sec:skyrmionMemory}, we give the moduli space formalism and results for the skyrmion dynamics.
Sec.~\ref{sec:conclusions} contains our concluding remarks.
In Appendix~\ref{sec:symmetry_arguments}, we explain the symmetry arguments that are used to simplify calculations in the main text.
In Appendix~\ref{sec:guidingCentre}, we give the exact treatment of the skyrmion trajectories, which turns out to be somewhat technical.

\section{The model for a synthetic antiferromagnet}
\label{sec:model}

\subsection{Energy}

We consider two ferromagnetic layers, the {\it upper} and the {\it lower} layer denoted by $U$ and $L$, respectively. In the continuum approximation, and assuming that the magnetization does not vary across the layer thickness, we define the magnetization $\Magn=\Magn^U, \Magn^L$ in the two layers, then the total energy functional is
\[
    E_{\rm tot} = E_\text{SL}(\Magn^U) + E_\text{SL}(\Magn^L) + E_\text{int}(\Magn^U,\Magn^L),
\]
with the same intralayer energy form in each layer
\[
\Energy_\text{SL}(\Magn) = \frac{A}{M_s^2} \int \p_\mu \Magn\cdot \p_\mu \Magn\, d^3x + \frac{\DM}{M_s^2} \int \e_\mu\cdot (\p_\mu\Magn\times\Magn)\, d^3x  + \frac{\Anisotropy}{M_s^2} \int (1 - M_3^2)\, d^3x - \mu_0 \int \bHext \cdot \Magn\,d^3x
\]
and the interlayer energy
\[
\Eint = \frac{\Dcoupling}{M_s^2} \int \Magn^U\cdot \Magn^L\,d^2x.
\]
We normalize the magnetization to the saturation magnetization $M_s$, $\magnT=\Magn^U/M_s$, $\magnB = \Magn^L/M_s$  and measure length in units of $\ldw=\sqrt{A/\Anisotropy}$.
The total energy of the system is then
\begin{equation} \label{eq:energy}
\Etotal = \Energy_{\text{SL}}(\magnT) + \Energy_{\text{SL}}(\magnB) + \Eint(\magnT,\magnB),
\end{equation}
with the normalized form of the single-layer energy
\begin{equation}  \label{eq:energy1}
\Energy_\text{SL}(\magn)= \frac{1}{2} \int \p_\mu \magn\cdot\p_\mu \magn\, d^2x
+ \dm \int \e_\mu\cdot (\p_\mu\magn\times\magn)\,d^2x
 + \frac{1}{2} \int \left[1- \left(\magn_3\right)^2 \right]\,d^2x
 - \int \bhext\cdot\magn\,d^2x
\end{equation}
and the interaction energy
\begin{equation}
\Eint (\magnT,\magnB)= \interJ \int \magnT\cdot \magnB\,d^2x
\end{equation}
in units of $2A t$, with $t$ the thickness of each layer, and the dimensionless parameters are expressed in terms of the material parameters as
\[
\dm = \frac{\DM}{2\sqrt{A\Anisotropy}},\qquad \interJ = \frac{\Dcoupling}{2\Anisotropy t},\qquad \bhext = \frac{\bHext}{H_0},\quad H_0 = \frac{2\Anisotropy}{\mu_0 M_s} 
\]

The equations of motion for the magnetization in the two layers are
\begin{equation} \label{eq:LLG}
\begin{split}
\p_\tau\magnT & = -\magnT\times(\heff(\magnT) - \interJ \magnB) + \alpha\,\magnT\times\p_\tau\magnT  \\
\p_\tau\magnB & = -\magnB\times(\heff(\magnB) - \interJ \magnT) + \alpha\,\magnB\times\p_\tau\magnB
\end{split}
\end{equation}
where we have included a Gilbert damping term with parameter $\alpha$ and the effective fields are
\[
\heff(\magn) = -\frac{\delta \Energy_{\text{SL}}(\magn)}{\delta \magn}= \Delta\magn - 2\dm (\e_\mu\times\p_\mu\magn) + m_3 \e_3 + \bhext.
\]
Time is measured in units of $\tau_0=M_s/(2\gamma K)$.

\begin{table}[h]
\centering
\begin{minipage}{0.45\linewidth}
\centering
\begin{tabular}{c c}
\hline\hline
\textbf{Parameter} & \textbf{Value} \\
\hline
$A$ &$1.50 \times 10^{-11}\,{\rm J/m}$\\
\Anisotropy &$0.60\times10^6\,{\rm J/m^3}$\\
\DM& $3.05\times 10^{-3}\,{\rm J/m^2}$\\
$A_c$&$4.80\times 10^{-5}\,{\rm J/m^2}$\\
$t$& $0.40\times 10^{-9}\,{\rm m}$\\
$M_s$& $0.58\times10^6\, {\rm A/m}$\\
\hline\hline
\end{tabular}
\end{minipage}
\hfill
\begin{minipage}{0.45\linewidth}
\centering
\begin{tabular}{c c}
\hline\hline
\textbf{Parameter} & \textbf{Value} \\
\hline
$\dm$ & $0.51$ \\
$\interJ$ & $0.10$ \\
\hline\hline
\textbf{Unit} & \textbf{Value} \\
\hline
$\mu_0 H_0$ & 2.07\,{\rm T} \\
$\ldw$ & $5.0\,{\rm nm}$ \\
$\tau_0$ & $2.75\,{\rm ps}$ \\  
\hline\hline
\end{tabular}
\end{minipage}
\caption{An example of values for physical parameters and the corresponding dimensionless parameters and units of space and time used in the present formulation. All values are taken from Ref.~\cite{PhysRevB.110.094430}.}
     \label{tab:units}
\end{table}

\subsection{Skyrmions in the bilayer}

Skyrmions are characterized by the topological number
\begin{equation} \label{eq:SkyrmionNumber}
\Skyrmion(\magn) = \frac{1}{4\pi} \int \skyrmion\, d^2x,\qquad \skyrmion = \magn\cdot(\p_2\magn\times\p_1\magn)
\end{equation}
where the integral is taken over an entire layer.
From the symmetries of the energy, we can assume $\magnT=-\magnB$. This then gives $\Skyrmion(\magnT)=-\Skyrmion(\magnB)=:\Skyrmion$.
We call the combined skyrmion configuration in $\magnT,\magnB$ a {\it bilayer skyrmion}.

\begin{figure}[t]
\begin{center}
\includegraphics[width=0.4\textwidth]{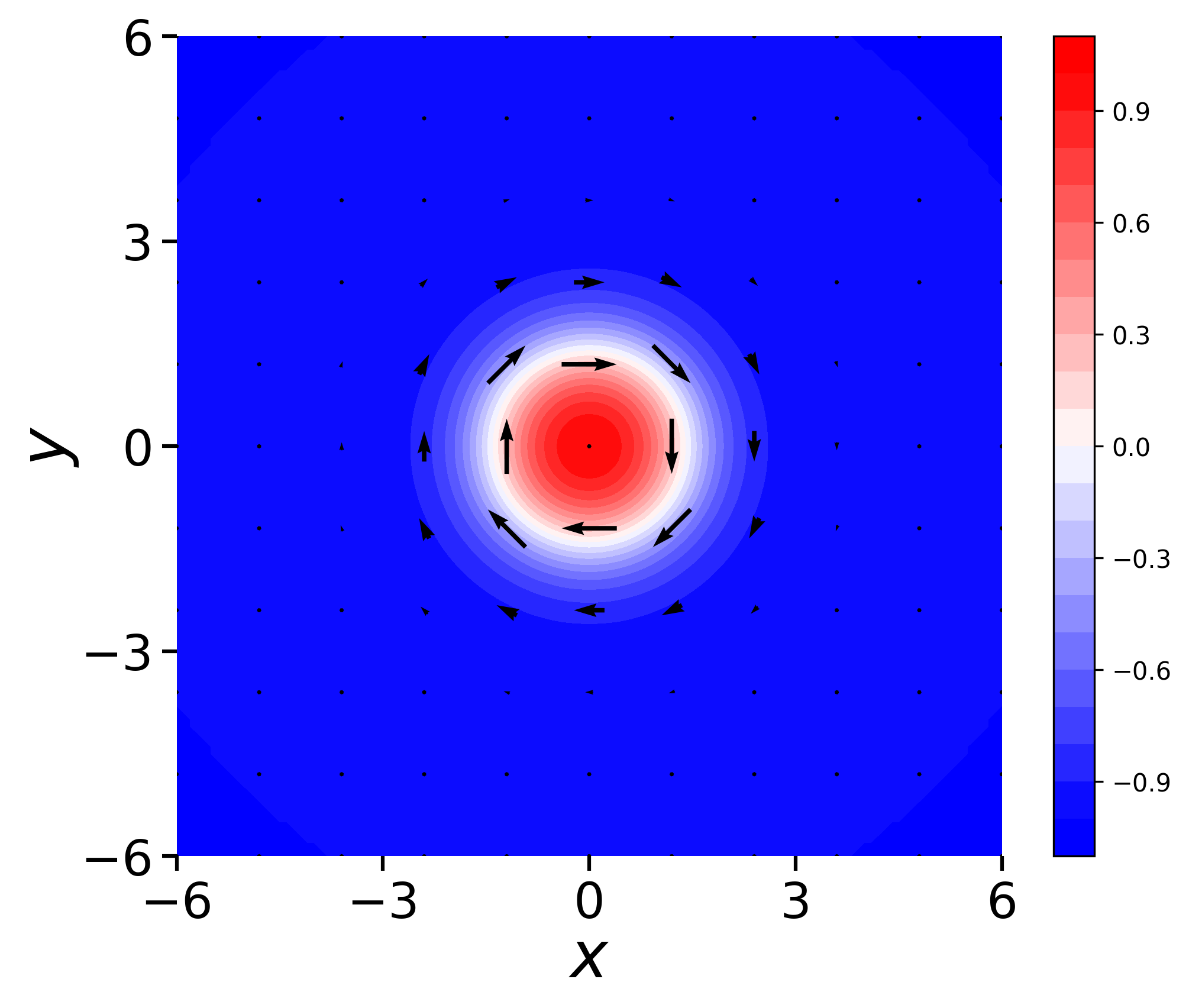}
\includegraphics[width=0.4\textwidth]{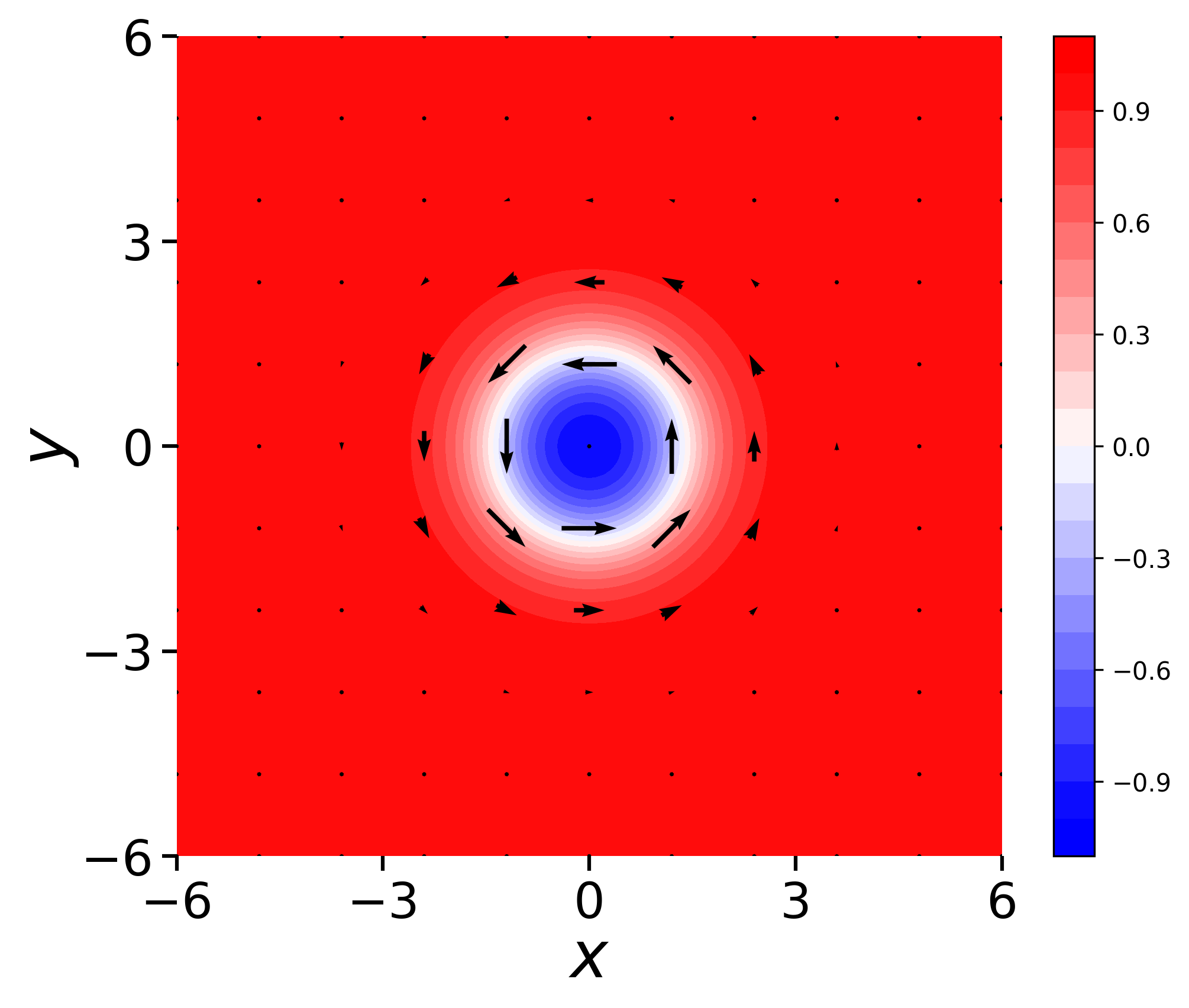}
\caption{
A bilayer skyrmion for parameter value $\dm=0.5$. 
(Left) The skyrmion magnetization $\magnB$ in the lower layer and (Right) the skyrmion magnetization $\magnT$ in the upper layer, where $\magnT = -\magnB$.
The skyrmion configurations are identical to skyrmions in a monolayer with the same intralayer parameter values.
Vectors show the projection of the magnetization on the $xy$ plane and the third component of the magnetization is shown via a colour code.
}
\label{fig:skyrmionBL}
\end{center}
\end{figure}

For a single layer, there is a static axisymmetric skyrmion solution in the parameter range $0 < \dm < 2/\pi$ \cite{Wang2018} when no external field is present, i.e. $\bhext=0$. The profile of the bilayer skyrmion is in fact the same as the skyrmion in a single layer with the same parameter $\dm$, independent of the interlayer coupling. This can be justified by a general argument.
The principle of symmetric criticality \cite{1979_CMP_Palais} guarantees that the minimum of the energy functional when we impose $\magnT=-\magnB$ is a stationary point of the full energy. 
When we substitute this ansatz into the energy, the interlayer coupling term becomes a constant and thus irrelevant to the Euler-Lagrange equations.
Therefore, the profile of this solution is identical to that of a single layer.
Note that this observation does not hold for other interlayer interactions such as the interlayer dipole interaction \cite{Bhukta2024,2017_NatComm_HrabecRohart}: to first approximation, that interaction could be modelled as $\int(\magnB)_3(\magnT)_3d^2x$. Applying the steps above would lead to an effective energy for the bilayer skyrmion profile which contains an easy-plane anisotropy term proportional to the dipole interaction strength, and the profile would thus depend on this parameter.

\section{Numerical simulation of Skyrmion under field gradient}
\label{sec:numericalSimulation}

\subsection{Numerical simulation}
\label{sec:field_gradient_simulation}

We perform numerical simulations of skyrmion dynamics applying a field gradient along the $x_1$ direction, in the form
\begin{equation}  \label{eq:fieldGradient}
\bhext = \hext(x_1)\,\e_3,\qquad \hext(x_1) = -g x_1\,e^{-(x_1/\ell)^2}
\end{equation}
where $g, \ell$ are constants.
This gives a pure gradient close to $x_1=0$, where the skyrmion is located, while the field decays at large distances, $|x_1|\to\infty$.
We will use $\ell=10$ in the simulations, while a typical skyrmion radius is $1.4$. The constant $g$ has units of $\frac{2K}{\mu_0 M_sl_w}$. For the parameters in Table \ref{tab:units}, this corresponds to $3.29\times 10^{14}\,{\rm A/m^2}$ or $0.41\,{\rm T/nm}$. 

As an initial condition, we use a bilayer skyrmion that is static in the absence of the field, and run the {\it conservative} algorithm, including the field gradient \eqref{eq:fieldGradient}.
It is well known that in this case a skyrmion in a single layer moves perpendicular to the direction of the gradient.

Within the conservative Landau-Lifshitz equation, under the approximation of a pure gradient $\bhext=-gx_1\,\e_3$, the skyrmion velocity is $\bm{\vel}=(0,\vel_2)$ with \cite{1991_NPB_PapanicolaouTomaras}
\begin{equation} \label{eq:velocity_pureGradient}
\vel_2 = -\frac{g\tmagn}{4\pi\Skyrmion},\qquad\tmagn = \int (m_3-m_0)\,d^2x
\end{equation}
where $\tmagn$ is the total magnetization in the perpendicular direction, and $m_0\e_3$ is the magnetization at spatial infinity.
The two layers have opposite magnetization and opposite skyrmion number.
Therefore, the velocities of the skyrmions in the two layers are expected to be equal according to formula \eqref{eq:velocity_pureGradient}.
Due to this, the skyrmions remain on top of each other while moving. In effect, the two layers are thus decoupled. Therefore when the field gradient is switched off, the skyrmions remain spontaneously pinned in their final position, as in the case of a single skyrmion \cite{1991_NPB_PapanicolaouTomaras}.
We have confirmed this picture numerically.

\begin{figure}[t]
\begin{center}
\includegraphics[width=0.4\textwidth]{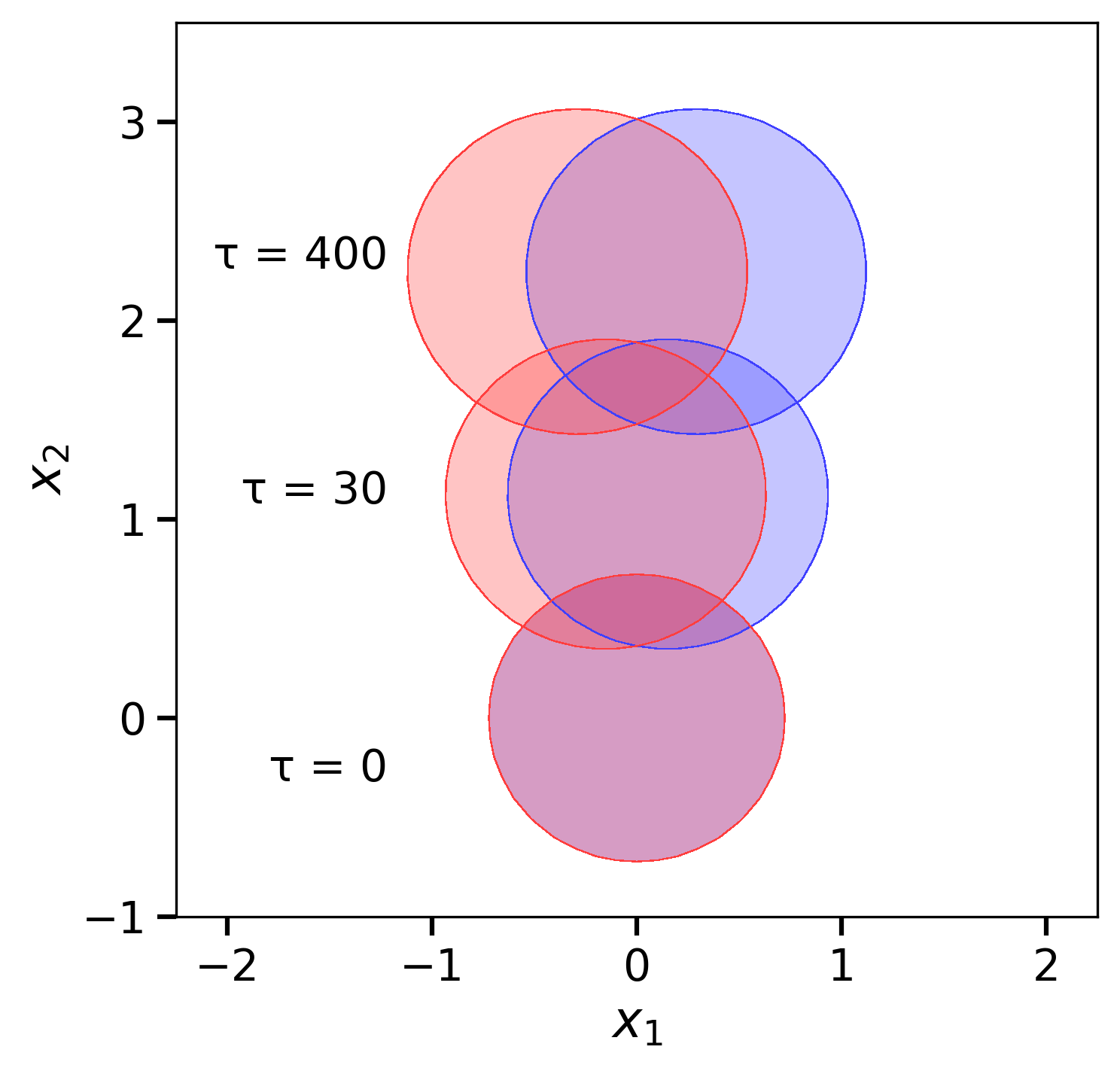}
\caption{Snapshots of the bilayer skyrmions propagating under a magnetic field gradient with $g=0.05$ (lower in red, upper in blue).
Shown are the contours for $L_3=0.75,\, U_3=-0.75$.
(Note that the skyrmion radius, measured at $U_3=0, L_3=0$, is 1.4.)
The simulation starts from the static skyrmion, for parameter value $\dm=0.5$ and damping parameter $\alpha=0.1$,
The field is switched on at time $\tau=0$.
}
\label{fig:skyrmionBL_gradientDamping}
\end{center}
\end{figure}

In the following set of simulations, we include damping and run the LLG equation \eqref{eq:LLG}.
Fig.~\ref{fig:skyrmionBL_gradientDamping} shows snapshots of the simulation.
Both skyrmions move primarily in the positive $x_2$ direction while, due to damping, one drifts to the right and the other to the left.
Initially, the skyrmions are driven by the field gradient.
As the skyrmions drift away from each other, the attracting force between them increases and acts in each skyrmion opposite to the force of the field gradient.
Therefore, as their separation increases, the net force decreases and the propagation velocity along $x_2$ is reduced.
For a not too strong field gradient, the attractive force will eventually balance the force due to the field gradient giving a zero net force.
At this point, the bilayer skyrmion motion will be halted with the two skyrmions remaining at a distance from each other.

However, the attractive force does not increase indefinitely: it reaches a maximum and then attenuates to zero at large separation. In specific cases, the force profile has been numerically investigated \cite{2017_srep_KoshibaeNagaosa}.
Thus, if the field gradient is sufficiently strong, the attractive force is never sufficient to balance the force due to field gradient and the skyrmions dissociate. Note that the point of maximum force and thus the maximum possible separation is independent of the interlayer ccoupling strenght $\interJ$.

When we switch off the field gradient at any time during motion, the attractive force between them makes the skyrmions reverse their direction of motion.
We observe the surprising phenomenon that the skyrmions eventually return to their original positions.

\begin{figure}[t]
\begin{center}
\includegraphics[width=0.4\textwidth]{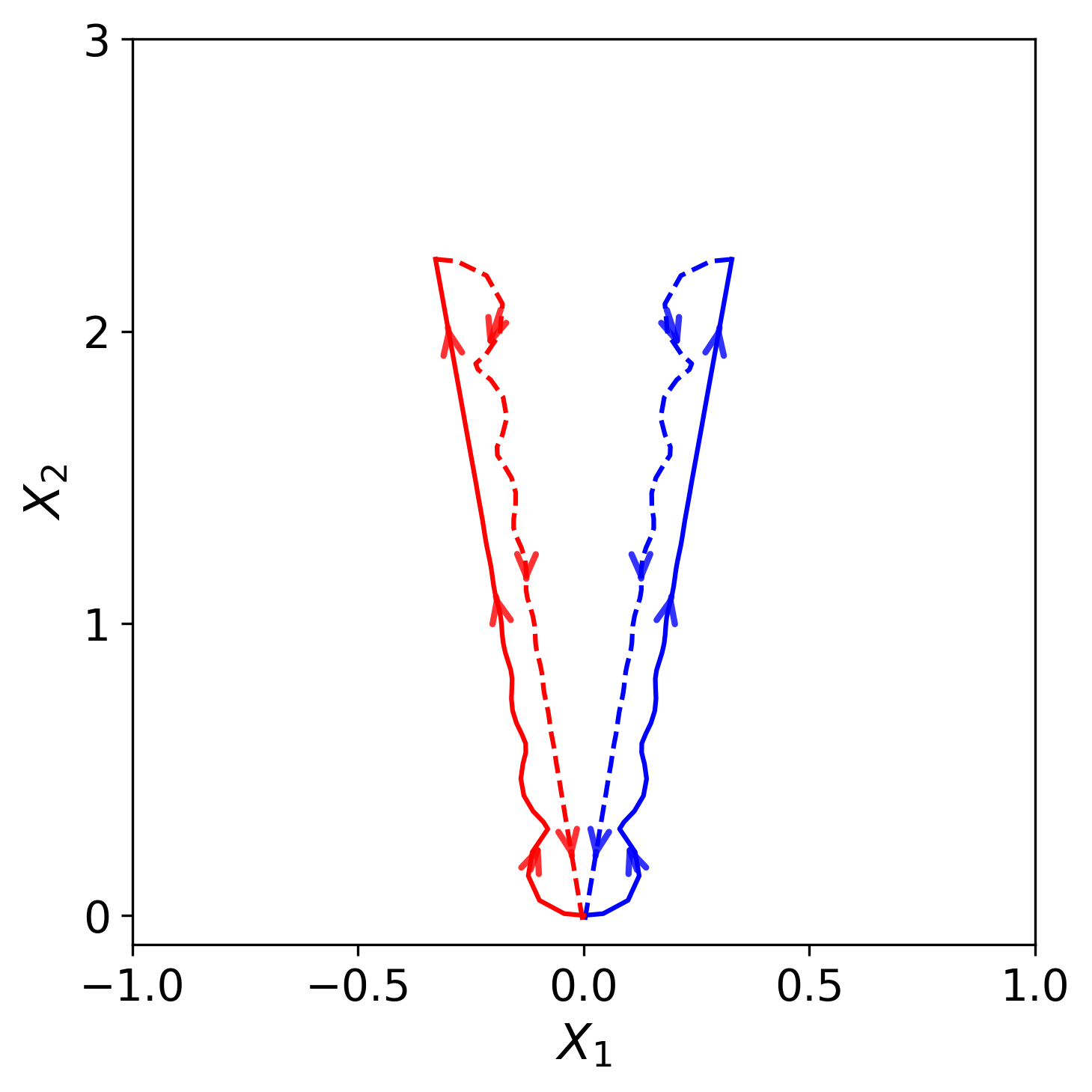}
\caption{Motion of the centers of the two skyrmions (lower in red, upper in blue) in the bilayer according to definition \eqref{eq:XY}.
The simulation is the same as in Fig.~\ref{fig:skyrmionBL_diffgradient}.
A field gradient is applied for times $0 < \tau < 400$ with $g=0.05$ and the field is switched off at $\tau=400$.
Solid lines represent motion under the applied field gradient, while dashed lines show the trajectories after the field is switched off.
}
\label{fig:XY_trajectory}
\end{center}
\end {figure}

For a quantification of the skyrmion motion, we follow their trajectories.
We trace the position of the skyrmions using normalized moments of magnetization,
\begin{equation} \label{eq:XY}
    X_1 = \frac{1}{\tmagn} \int x_1 (m_3-m_0)\,d^2x,\qquad 
    X_2 = \frac{1}{\tmagn} \int x_2 (m_3-m_0)\,d^2x.
\end{equation}
This is a standard measure for skyrmion position encoded as a function in MuMax$^3$ \cite{Joos_2023_AppliedPhys}. 
The formulae are applied separately in each layer with $m_0=1$ in the upper layer and $m_0=-1$ in the lower layer.
Fig.~\ref{fig:XY_trajectory} shows the trajectories of the two skyrmions under a field gradient with solid lines on the outward trajectory and dashed lines on the return trajectory.
It seems mysterious that, although the $(X_1, X_2)$ trajectories forward and back are different, the skyrmions return to their original positions. 

To explain this, it is illuminating to instead use the normalized moments of the topological density to track skyrmion position:
\begin{equation} \label{eq:guidingCenter}
    \guidex = \frac{1}{4\pi\Skyrmion} \int x_1\skyrmion\,d^2x,\qquad
    \guidey = \frac{1}{4\pi\Skyrmion} \int x_2\skyrmion\,d^2x.
\end{equation}
The measure of skyrmion position $(\guidex,\guidey)$ is called the {\it guiding center} and it has the fundamental property that, within the conservative model, it remains invariant when no forces act on the skyrmions \cite{1991_NPB_PapanicolaouTomaras}.

\begin{figure}[h]
\begin{center}
\includegraphics[width=0.4\textwidth]{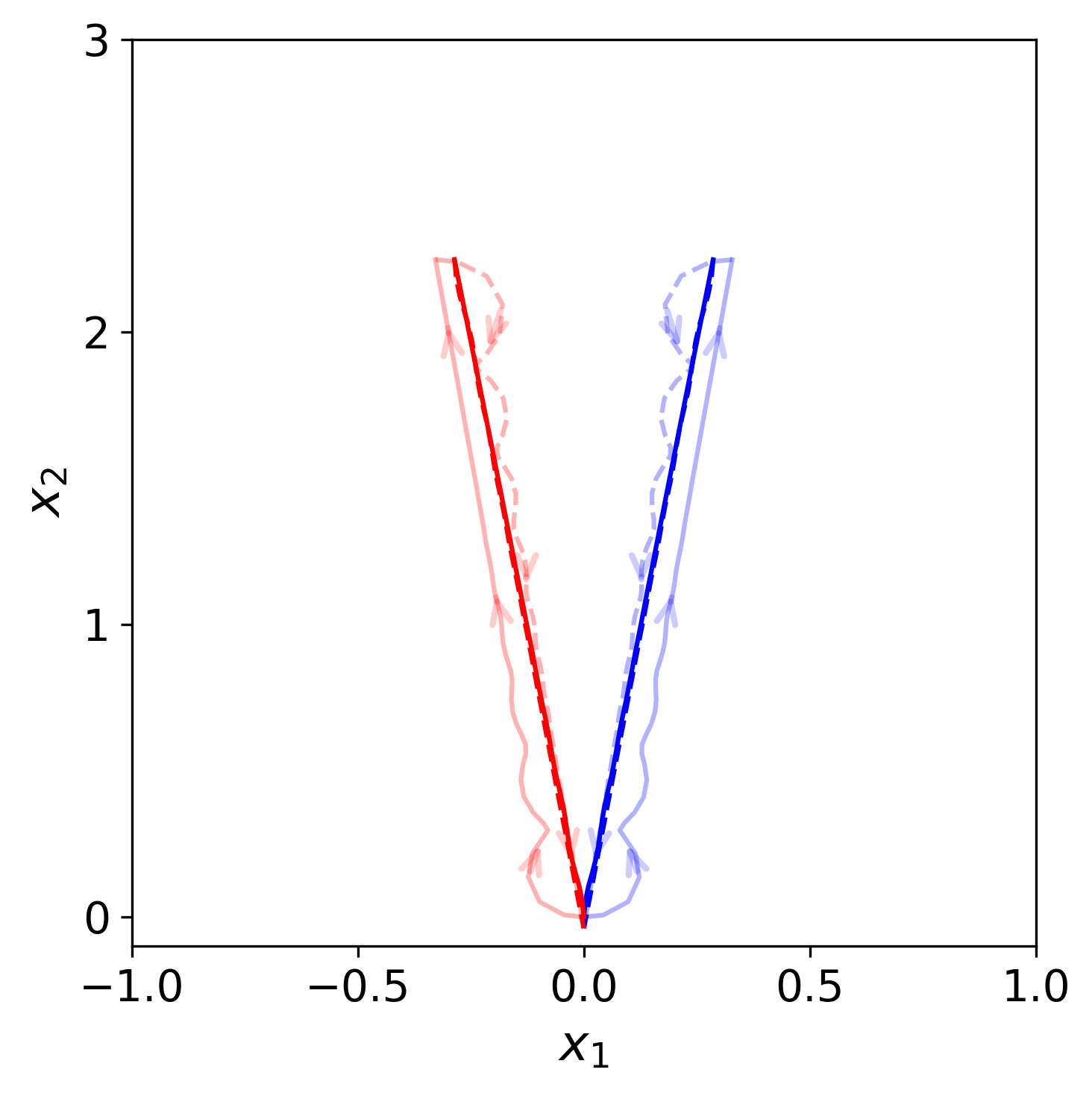}
\caption{The trajectories of the two skyrmions for the same simulation as in Figs.~\ref{fig:skyrmionBL_gradientDamping},\ref{fig:XY_trajectory}.
The trajectories for the lower and upper layer skyrmion are shown with red and blue lines, respectively.
Solid lines represent motion under the applied field gradient, while dashed lines show the trajectories after the field is switched off.
The trajectories of the guiding centres $(\guidex, \guidey)$ are shown by strong lines.
The faded lines show the trajectories for $(X_1, X_2)$ which were also presented in Fig.~\ref{fig:XY_trajectory} and are drawn here for comparison.
}
\label{fig:RxRyXY_trajectory}
\end{center}
\end{figure}

Fig.~\ref{fig:RxRyXY_trajectory} shows the trajectories of the skyrmions using $(\guidex, \guidey)$ and $(X_1, X_2)$ for the same simulation.
When the field is applied, the skyrmions eventually reach final positions which are different when measured via the two methods.
This is because the skyrmions are not axisymmetric when they are separated.
When the field is switched on, the guiding centers of the two skyrmions $(\guidex, \guidey)$ follow paths that are close to straight lines.
When the field is switched off, the guiding centers return to their initial positions, approximately retracing their spatial trajectories on the way forward.
This should be contrasted with the more complicated paths for $(X_1,X_2)$.
The trajectory of the guiding center gives a hint for the theoretical description of the phenomenon that we will present in the next sections.

\begin{figure}[t]
\begin{center}
\includegraphics[width=0.4\textwidth]{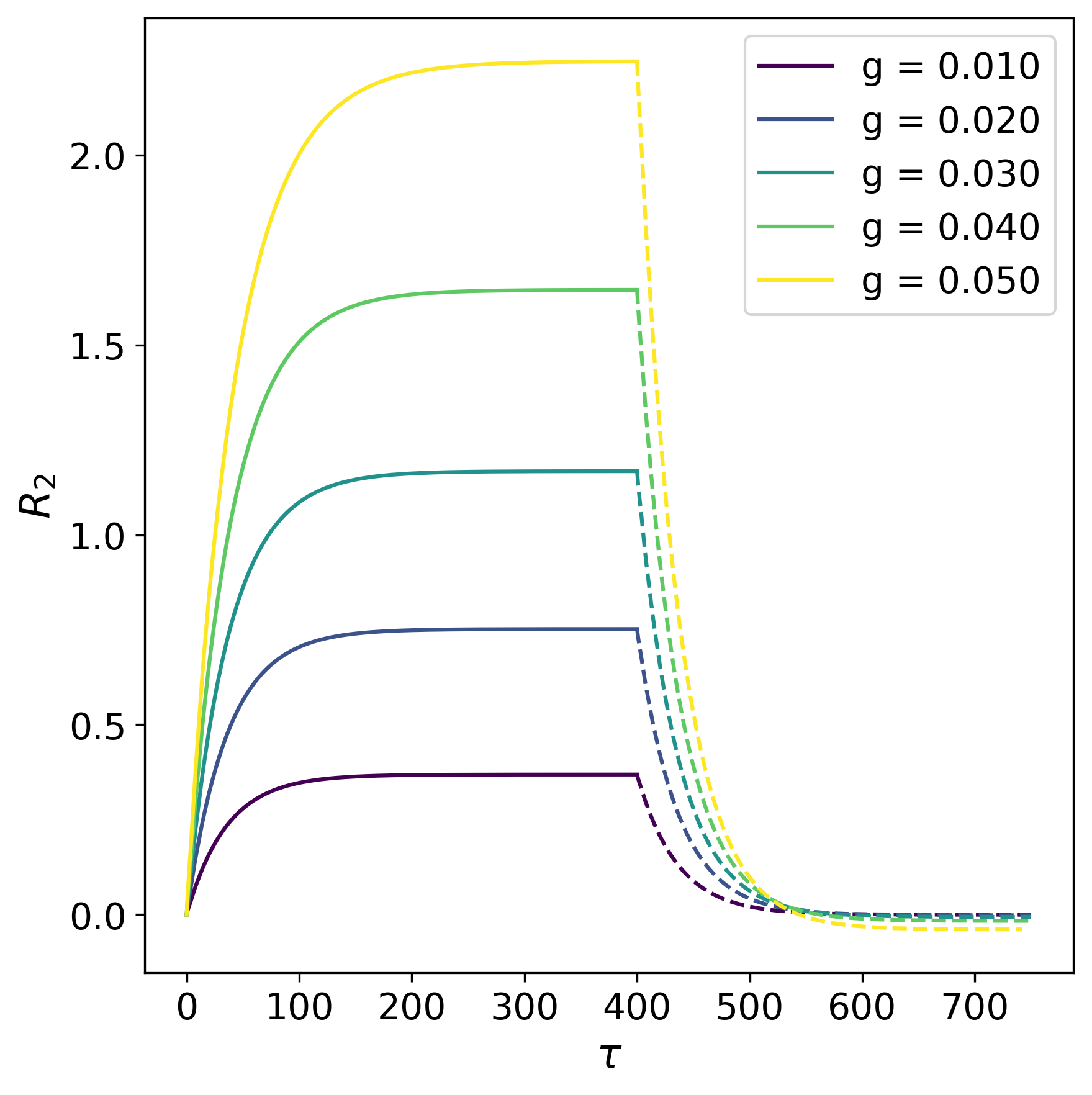}
\caption{The guiding centre component $\guidey$ for skyrmions in both layers as a function of time under various field gradient strengths $g$.
The field gradient is switched on at $\tau = 0$ and switched off at $\tau = 400$.
}
\label{fig:skyrmionBL_diffgradient}
\end{center}
\end{figure}

\begin{figure}[t]
\begin{center}
\includegraphics[width=0.4\textwidth]{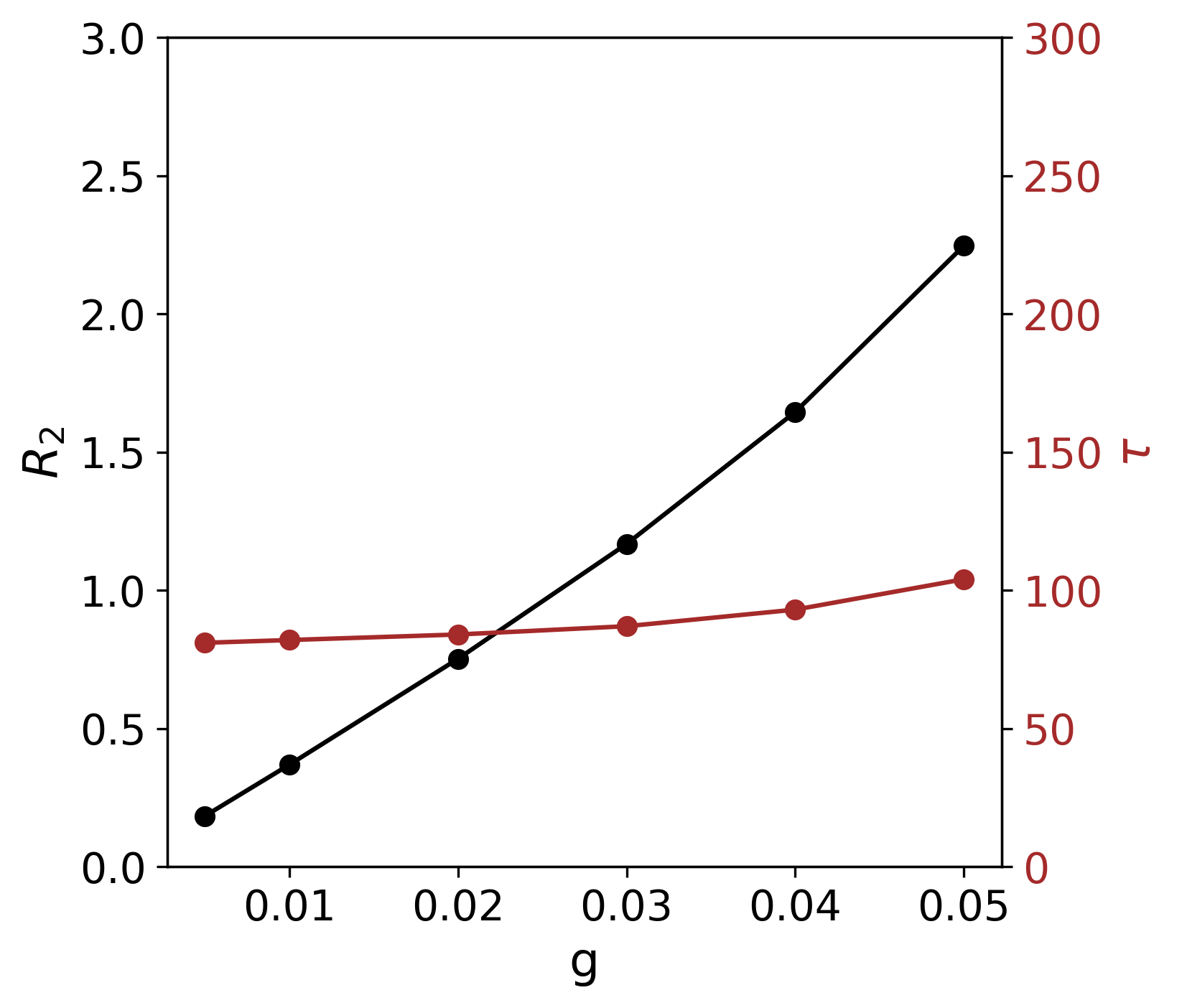}
\caption{The black line shows the terminal position of the skyrmion measured by the guiding centre coordinate $\guidey$, for applied field gradient with different strengths $g$.
The brown line shows the time taken to reach $90\%$ of the maximum value of $\guidey$.
}
\label{fig:maximum-skyrmion-displacement}
\end{center}
\end{figure}

Besides the spatial trajectory we are also interested in the time it takes the skyrmions to reach different positions. We run a series of simulations using different field gradient strengths $g$.
The field is applied from $\tau=0$ to $\tau = 400$, at which point the skyrmions have reached their terminal position and they have stopped moving, and is then switched off.
Fig.~\ref{fig:skyrmionBL_diffgradient} shows the $\guidey$ component as a function of time (it is identical for the two skyrmions).
The skyrmions start with a large velocity $d\guidey/d\tau$, which is almost linearly increasing with $g$, and this is gradually reduced as the interaction force between the skyrmions balances the force due to the field.
Fig.~\ref{fig:maximum-skyrmion-displacement} shows that the terminal $\guidey$ is an increasing function of the gradient strength $g$.
On the other hand, the same figure shows that the time it takes to reach the final $\guidey$ is a much less sensitive function of $g$.

When the field is switched off, the skyrmions move initially fast in the reverse direction, and the velocity is gradually reduced until they stop.
Fig.~\ref{fig:skyrmionBL_diffgradient} shows that $\guidey$ returns to its original value $\guidey=0$, but the velocity on the way back is far from a retracing of the velocity of the forward motion.

\section{Skyrmion positional memory}
\label{sec:skyrmionMemory}

\subsection{Thiele equation and skyrmion equilibrium}
\label{sec:virial}

We start the theoretical study of the dynamics of skyrmions by studying the Thiele equations, as in \cite{2020_JMMM_Zhou}.
For a skyrmion in a single layer (or in a film) propagating at a constant velocity $(\vel_1, \vel_2)$ under an external field, the Thiele equation is \cite{1973_PRL_Thiele}
\begin{equation*}  \label{eq:virial1}
\vel_\nu (\emn\,\Skyrmion + \alpha\, \Gamma_{\mu\nu}) = F_\mu
\end{equation*}
where
\begin{equation*}  \label{eq:integratedQuantities1}
F_\mu = \frac{1}{4\pi} \int \magn\cdot \p_\mu\bhext\,d^2x,\qquad
\Gamma_{\mu\nu} = \frac{1}{4\pi}\int \p_\mu \magn\cdot \p_\nu \magn\,d^2x.
\end{equation*}
In the bilayer system, we have a similar equation for each layer,
\begin{equation} \label{eq:virial_field}
\begin{split}
& \vel_\nu^U (\Skyrmion \emn + \alpha \Gamma_{\mu\nu}^U) =  F^U_\mu \\
& \vel_\nu^L (-\Skyrmion \emn + \alpha \Gamma_{\mu\nu}^L) = F^L_\mu,
\end{split}
\end{equation}
where 
\begin{equation} \label{eq:integratedQuantities}
F^U_\mu = \frac{1}{4\pi} \int \magnT\cdot \p_\mu\bhext\,d^2x + \frac{\interJ}{4\pi} \int \magnB\cdot\partial_\mu \magnT\,d^2x,\qquad
\Gamma_{\mu\nu}^U = \frac{1}{4\pi}\int \p_\mu \magnT\cdot \p_\nu \magnT\,d^2x,
\end{equation}
with corresponding expressions for $F^L_\mu,\, \Gamma_{\mu\nu}^L$. The forces $F_\mu^{U,L}$ have two parts, the external force from the magnetic field gradient and the inter-skyrmion force coming from the interlayer coupling.

The symmetries of the system can be applied to simplify Eqs.~\eqref{eq:virial_field}. Under the assumption that the solution retains some symmetries of the energy, we find in Appendix \ref{sec:symmetry_arguments}:
\begin{equation} \label{eq:F_hez}
F_1 := F_1^U = -F_1^L,\,\qquad 
F_2^{U,L} = 0.
\end{equation}
This shows that both the external force and the inter-skyrmion force are directed along the field gradient. 
Similarly, we find 
\[
\Gamma^U_{\mu\nu}=\Gamma^L_{\mu\nu}=:\Gamma_{\mu\nu} = \text{diag}(\Gamma + \epsilon,\Gamma-\epsilon)_{\mu\nu}.
\]
The average eigenvalue $\Gamma$ gives the overall size of the matrix, while $\epsilon$ is due to the distortion of the skyrmion.
For simplicity we assume that the distortion is small, $\epsilon \ll \Gamma$, although the results do not depend on this.
Using the above, Eqs.~\eqref{eq:virial_field} simplify to
\begin{equation} \label{eq:virial_field_with_symmetries}
\begin{split}
& \vel_\nu^U (\Skyrmion \emn + \alpha \Gamma\delta_{\mu\nu}) =  F_\mu \\
& \vel_\nu^L (\Skyrmion \emn - \alpha \Gamma \delta_{\mu\nu}) = F_\mu.
\end{split}
\end{equation}

While the force due to the field gradient depends only weakly on the position of each skyrmion, the inter-skyrmion force depends strongly on the distance between them.
This indicates that a balance of forces may be achieved when
\begin{equation}
F_1 = 0 
\implies
    \int\magnT\cdot\p_1 (\bhext - \interJ\magnB) \,d^2x = 0,
\end{equation}
where we have used integration by parts.
The solution of the integral equation would give the equilibrium separation between the skyrmions.
If the external force is larger than the maximum possible inter-skyrmion force, then no such solution is possible.

\subsection{Skyrmion trajectory and memory} 
\label{sec:general_Thiele_eqns}

Now we investigate the transient behaviour when the field gradient is turned on or off.
We assume that over a short timescale there is a solution with approximately constant velocity, but over a larger timescale the velocity can vary, the profile of the skyrmion adjusting instantaneously as it does so. 
Under the above assumption, the virial relations \eqref{eq:virial_field_with_symmetries} are valid at every instant and we may replace the constant $v_\mu$ with a derivative $\dot{x}_\mu$, giving a set of coupled differential equations,
\begin{equation} \label{eq:Thiele}
\begin{split}
& \dot{x}_\nu^U (\Skyrmion \emn + \alpha \Gamma \delta_{\mu\nu}) =  F_\mu \\
& \dot{x}_\nu^L (\Skyrmion \emn - \alpha \Gamma \delta_{\mu\nu}) = F_\mu.
\end{split}
\end{equation}
It is now useful to define new variables: the relative position of the skyrmion centres $(d_1,d_2)$ and their centre of mass $(D_1,D_2)$, respectively,
\[
d_\mu = \frac{1}{2}(x_\mu^U - x_\mu^L),\qquad
D_\mu = \frac{1}{2}(x_\mu^U + x_\mu^L).
\]
By adding and subtracting Eqs.~\eqref{eq:Thiele}, we obtain 
\begin{equation} \label{eq:Xx_virial}
\begin{split}
& \Skyrmion \emn\dot{d}_\nu + \alpha\Gamma\dot{D}_\mu = 0 \\
& \Skyrmion \emn\dot{D}_\nu + \alpha\Gamma\dot{d}_\mu= F_\mu.
\end{split}
\end{equation}
The first equation gives
\begin{equation} \label{eq:Xdot_propto_xdot}
\dot{D}_\mu = -\frac{Q}{\alpha\Gamma} \emn \dot{d}_\nu \implies  D_\mu = D_\mu(0) - \frac{Q}{\alpha\Gamma}\emn d_\nu.
\end{equation}
The result shows that the skyrmions move on either side of the $x_2$ axis at an angle $\theta$ given by
\begin{equation} \label{eq:angle_of_motion}
\tan\theta = \alpha\Gamma.
\end{equation}
Equations~\eqref{eq:Xdot_propto_xdot}, \eqref{eq:angle_of_motion} show that, assuming $\Gamma$ is constant, the direction of motion of skyrmions is constant despite the fact that the external force from the field gradient and the inter-skyrmion force may vary as the skyrmions move apart.

We further substitute Eq.~\eqref{eq:Xdot_propto_xdot} in the second of Eqs.~\eqref{eq:Xx_virial} and obtain
\begin{equation} \label{eq:dotx}
 \dot{d}_\mu = \frac{\alpha\Gamma}{\Skyrmion^2 + \alpha^2\Gamma^2}\, F_\mu.
\end{equation}
Since $(D_1,D_2)$ is given in terms of $(d_1,d_2)$ by Eq.~\eqref{eq:Xdot_propto_xdot}, then Eq.~\eqref{eq:dotx} can be thought of as an equation for $(d_1,d_2)$ only.

Since the force is directed along the axis between the skyrmions ($F_2=0$), the separation is in the same direction, that is, $d_2=0$.
Equations \eqref{eq:Xdot_propto_xdot}, \eqref{eq:dotx} give
\begin{equation} \label{eq:dotx1_X2}
    \dot{d}_1 = \frac{\alpha\Gamma}{\Skyrmion^2 + \alpha^2\Gamma^2}\, F_1(d_1),\qquad
    D_2 = D_2(0) + \frac{\Skyrmion}{\alpha\Gamma} d_1.
\end{equation}
Since $F_1$, $\Gamma$ are positive, $\dot{d}_1$ is positive: this means the skyrmion in the upper layer moves to the right and the skyrmion in the lower layer moves to the left, while both also move upwards since $\Skyrmion=1$.
The total force $F_1$ decreases as a function of $d_1$, so the velocity is initially at its fastest and decreases as the new equilibrium position of $d_1$ and $D_2$ is reached, when $F_1=0$.

Equations \eqref{eq:dotx1_X2} are also valid when we start with the skyrmions separated in this way and then remove the magnetic field, if now $F_1$ contains only the attractive inter-skyrmion force.
In this case, $F_1=0$ at $d_1=0$ and thus the separation is driven back to this point.
Because the angle of motion is independent of the force $F_1$, the motion will occur at the same fixed angle \eqref{eq:angle_of_motion}, therefore $D_2$ must also return to its initial value. We can also go further: even if $\Gamma$ changes as a function of $d_1$ (reflecting a change in size of the skyrmion, which is observed numerically), then the angle of propagation will change over the motion, but nevertheless the skyrmions will retrace their trajectories in space and thus return to their initial point.
Notably, the skyrmions can become elliptical or change size as a function of position without affecting the conclusion.

This theoretical analysis therefore reproduces the behaviour of bilayer skyrmions seen numerically in Sec.~\ref{sec:field_gradient_simulation}.
It also gives a prediction for the angle of motion of the skyrmions. These results are based on the approximation that the skyrmion profile instantaneously adjusts to a quasi-steady state configuration that does not depend on velocity.
This assumption cannot be exactly true: at the moments the field is switched on or off, there is a moment of irreducibly transient behaviour before the quasi-steady-state transient behaviour we describe here.

We can investigate this regime with an exact dynamical relation, finding that the initial angle can be different from the quasi-steady-state angle of propagation. This is done in Appendix~\ref{sec:guidingCentre}. The difference from the idealised behaviour above is most pronounced when the skyrmion is large. We can see in Fig.~\ref{fig:RxRyXY_trajectory} that the difference is small for the moderately sized skyrmions we consider here.

\section{Concluding remarks}
\label{sec:conclusions}

We studied the response of a skyrmion in a SAF bilayer under an external magnetic field gradient. We think of the bilayer skyrmion as made up of one skyrmion in each of the two layers.
The field gradient acts oppositely in the two layers, leading to skyrmion dynamics that is very different from the dynamics probed in the same system by electric currents.
The net skyrmion motion is perpendicular to the force due to the field gradient, even when damping is included.
Furthermore, the skyrmions appear to have memory of their original position in that they return to it after the field is switched off. 
This unusual behaviour is reproduced within a moduli space approximation.
The intuitive understanding is based on the result that the individual skyrmions move on almost straight lines when their position is tracked by their guiding centers \eqref{eq:guidingCenter}, at an angle that is almost independent of the speed they move or the force they experience. This means they move along the same trajectory both when the field is applied and after it is switched off.
The velocity of propagation and the maximum displacement are increasing, almost linear functions of the gradient strength, while the time to reach equilibrium is much less sensitive to changes in the gradient strength and it has a finite value even as the gradient strength goes to zero.

The reversing dynamical behaviour seen here is different from what can be observed in either ferromagnets or antiferromagnets. Ferromganetic skyrmions typically stay pinned in place at the new position when a force is removed; antiferromagnetic skyrmions typically continue in the same direction for some time before their kinetic energy is dissipated by damping.
This shows that the synthetic antiferromagnet does not always behave like an antiferromagnet and it should be studied in its own right.
The phenomenon of precise perpendicular confined motion is relevant to applications involving controlling the motion of skyrmions, such as racetrack memory.
Positional memory could be useful in its own right for various unconventional computing methods, such as token-based or neuromorphic computing.

\acknowledgments
The work of B.~B.-S. is supported by the ERC starting grant SINGinGR (Grant No.~101078061), under the European Union's Horizon Europe program for research and innovation.
The work of A.N. and S.K. is supported by the TOPOCOM project, which is funded by the European Union’s Horizon Europe Programme Horizon.1.2 under the Marie Skłodowska-Curie Actions (MSCA), Grant Agreement No. 101119608.
This work has greatly benefited from networking activities of the COST Action Polytopo, CA23134, supported by COST (European Cooperation for Science and Technology).

\appendix
\section{Constraining dynamical equations using symmetry\label{sec:symmetry_arguments}}

\subsection{Symmetries of the energy}

In the absence of magnetic field gradient, the intralayer energies $\Energy_{U,L}$ each have an $O(2)$ symmetry group \cite{2023_Thesis_Barton-Singer}, which can be generated from rotation in real and spin space simultaneously, and a simultaneous reflection in real and spin space:
\begin{equation} \label{eq:P2}
\magn(x)\mapsto P_2\magn(x) = P_2^{m} P_2^{x} \magn = 
\begin{pmatrix}
-m_1\\
m_2\\
m_3
\end{pmatrix}(x_1,-x_2).
\end{equation}
This reflection is so labelled because it inverts the $x_2$ direction in real space: reflections about any axis in real space with a corresponding reflection in spin space can be generated by combining with rotations. Skyrmions retain this symmetry. In the absence of any external magnetic field, they also have the inversion symmetry $\magn\mapsto-\magn$. The interlayer coupling has the interchange symmetry
\begin{equation}
    \magnT\leftrightarrow\magnB,
\end{equation}
and the presence of the interlayer coupling means that symmetries of the full energy must be symmetries of the intralayer energies acting simultaneously on $\magnT$ and $\magnB$, plus this interchange symmetry.

When we add the field gradient, we break the separate $O(2)$ symmetry group and interchange symmetry to get just $P_2$ above, and the analogously labelled $P_1$:
\begin{equation} \label{eq:P1}
\magn(x)\mapsto P_1\magn(x) = P_1^{m} P_1^{x} \magn = 
\begin{pmatrix}
m_1\\
-m_2\\
m_3
\end{pmatrix}(-x_1,x_2).
\end{equation}
combined with the inversion symmetry $\magn\mapsto-\magn$. For this to be a symmetry of the interlayer coupling also, it must be combined with the interchange symmetry:
\begin{equation} \label{eq:P1bilayer0}
\magnT\mapsto - P_1\magnB,\quad
\magnB\mapsto - P_1\magnT
\end{equation}
giving a symmetry of the total energy \eqref{eq:energy}.
We conclude that the symmetries for the bilayer system are $P_2$ in Eq.~\eqref{eq:P2} acting simultaneously on $\magnT$ and $\magnB$, and the ``combined $P_1$'' symmetry \eqref{eq:P1bilayer0}.

\subsection{Constraints on the dynamical equations}

We will assume that the traveling wave solution retains the symmetries of the energy:
\begin{equation}
    \magnT=P_2\magnT = -P_1\magnB,\qquad \magnB = P_2\magnB = -P_1\magnT.
\end{equation}
This in particular implies $\heff(\magnT)=P_2\heff(\magnT)=-P_1\heff(\magnB)$ etc.
We will use the symmetries to constrain the quantities $\Gamma_{\mu\nu},\,F_\mu,\, \Gamma_\mu$ appearing in the dynamical equations  \eqref{eq:virial_field}, \eqref{eq:dotR}.

For a magnetic field $\bhext=\hext(x)\e_3$, such as in Eq.~\eqref{eq:fieldGradient}, the $P_2$ symmetry gives $F_2^{U,L}=-F_2^{U,L}$, and the combined $P_1$ symmetry \eqref{eq:P1bilayer0} gives $F_\mu^U=-F_\mu^L=:F_\mu$.
Thus
\begin{equation} 
F_1^U = -F_1^L =: F_1, \,\qquad F_2^U=F_2^L = 0.
\end{equation}
Both internal and external forces are antisymmetric, but the interpretation is different: the force due to the field gradient represents an external force pushing the skyrmions in opposite directions due to their opposite magnetisation, while the interlayer force is an interaction force pulling the skyrmions back together, acting equally and oppositely in a manner akin to Newton's third law.
$F_1$ is non-zero and the explicit integral is presented in the main text as Eq.~\eqref{eq:F_hez}.

We can apply the same method to $\Gamma_{\mu\nu}$. Symmetry under $P_2$ gives $\Gamma^{U,L}_{12}=\Gamma^{U,L}_{21} = 0$ and symmetry \eqref{eq:P1bilayer0} tells us that $\Gamma^U_{\mu\nu}=\Gamma^L_{\mu\nu}$.
We conclude that
\begin{equation}
    \Gamma^U_{11} =\Gamma^L_{11}=:\Gamma_{11},\quad \Gamma^U_{22}=\Gamma^L_{22}=:\Gamma_{22},\qquad\Gamma^{U,L}_{12}=\Gamma^{U,L}_{21} = 0.
\end{equation}

To constrain $\Gamma_\mu$ from symmetry, we need the pseudotensor property of the antisymmetric tensor:
\begin{equation*}
\epsilon_{ijk}M_{ip}M_{jq}M_{kr} = \det(M)\epsilon_{pqr},
\end{equation*}
which implies in particular that considering a general reflection in spin space $P$ (which has the properties $P^2 =I$, $\det P=-1$), 
\begin{equation}
\label{eq:P_identity}
\epsilon_{ijk}u_i (Pv)_j (Pw)_k = - \epsilon_{ijk}(Pu)_i v_j w_k.
\end{equation}
We use this identity when we apply the $P_2$ symmetry to $\Gamma_\mu^U$:
\begin{align*}
\begin{split}
\Gamma^U_\mu(\magnT,\magnB) &=\Gamma_\mu^U(P_2\magnT,P_2\magnB) = \int (\heff(\magnT)+\bm{h}+j_cP_2\magnB)\cdot(P_2\magnT\times \partial_\mu P_2\magnT)\\
&=-\int (P_2\heff(\magnT)+P_2\bm{h}+j_c\magnB)\cdot(\magnT\times(\partial_1,-\partial_2) \magnT)\\
&=-\int (\heff(\magnT)+\bm{h}+j_c\magnB)\cdot(\magnT\times(\partial_1,-\partial_2) \magnT).
\end{split}
\end{align*}
Thus $\Gamma^U_1=-\Gamma^U_1$, and we similarly find $\Gamma^L_1=-\Gamma^L_1$. We similarly use the combined $P_1$ symmetry \eqref{eq:P1bilayer0} and the identity \eqref{eq:P_identity} to find $\Gamma_{1,2}^U=\Gamma_{1,2}^L$. Thus
\begin{equation}
\Gamma_1^U=\Gamma_1^L=0,\,\quad     \Gamma_2^U = \Gamma_2^L=:\Gamma_2.
\end{equation}

\section{Guiding centre dynamics}
\label{sec:guidingCentre}
\subsection{Exact transient behaviour}
We may describe the skyrmion motion without the adiabatic approximation, by starting from the fundamental dynamical equation for the topological density \cite{1991_NPB_PapanicolaouTomaras},
\begin{equation}
 \dot{\skyrmion} = -\emn\,\p_\mu (\bm{g}\cdot\p_\nu\magn),\qquad
 \bm{g} = \alpha_1 (\heff + \alpha\,\magn\times\heff)
\end{equation}
where $\alpha_1=1/(1+\alpha^2)$, and $\heff=-\frac{\delta E}{\delta \magn}$ is the effective field in the Landau-Lifshitz equation.
Note that the component of $\heff$ parallel to $\magn$ does not enter the equations, so we can take $\bm{f}\perp\magn$ for simplicity. The field $\bm{f}=\bm{0}$ for a static configuration, so we can think of it as measuring distortions away from a static solution.

We start by taking the time derivative of the guiding center \eqref{eq:guidingCenter},
\begin{equation} \label{eq:fundamental_dynamics}
\begin{split}
\dot{R}_\mu & = \frac{1}{4\pi\Skyrmion}\int x_\mu \dot{\skyrmion}\, d^2x 
= -\frac{\epsilon_{\lambda\nu}}{4\pi\Skyrmion} \int x_\mu\, \p_\lambda (\bm{g}\cdot\p_\nu\magn)\, d^2x
= \frac{\emn}{4\pi\Skyrmion} \int (\bm{g}\cdot\p_\nu\magn)\, d^2x  \\
 & = \frac{\alpha_1 \emn}{4\pi\Skyrmion} \left[ \int \heff\cdot\p_\nu\magn\,d^2x
  + \alpha \int (\magn\times\heff) \cdot\p_\nu\magn\, d^2x \right]
\end{split}
\end{equation}
where a partial integration has been performed and the boundary term was assumed to vanish.

An external field is readily taken into account by letting $\heff \to \heff + \bhext$.
For the bilayer, we write an equation of the form \eqref{eq:fundamental_dynamics} for each layer and include the interlayer interaction by substituting $\frac{\delta E}{\delta \magnT}=\heff(\magnT) + \bhext - \interJ \magnB$ and $\frac{\delta E}{\delta \magnB}=\heff(\magnB) + \bhext - \interJ \magnT$.
We obtain the system
\begin{equation} \label{eq:dotR}
\begin{split}
 \Skyrmion \dot{R}_\mu^U & = \alpha_1 \emn (F^U_\nu - \alpha \Gamma^U_\nu) \\
 -\Skyrmion \dot{R}_\mu^L & = \alpha_1 \emn (F^L_\nu - \alpha \Gamma^L_\nu)
\end{split}
\end{equation}
where
\begin{equation} \label{eq:F-G}
F_\mu^U = \frac{1}{4\pi} \int (\bhext - \interJ\magnB) \cdot \p_\mu\magnT\,d^2x,\qquad
\Gamma_\mu^U = \frac{1}{4\pi} \int (\heff(\magnT) + \bhext - \interJ \magnB)\cdot(\magnT\times\p_\mu\magnT)\, d^2x
\end{equation}
with corresponding expressions for $F_\mu^L,\, \Gamma_\mu^L$.

The expressions in Eq.~\eqref{eq:F-G} include an ``internal'' part containing $\heff$, an external field part, and an interlayer interaction part. The term $\int \heff\cdot\p_\mu\magn\,d^2x$ vanishes and has been omitted in $F_\mu$ since $\heff$ comes from an energy functional with translational invariance \cite{1973_PRL_Thiele,1991_NPB_PapanicolaouTomaras}.
While Eqs.~\eqref{eq:dotR} appear similar to the equations of motion in Sec.~\ref{sec:general_Thiele_eqns}, the present relations are exact and do not rely on any assumptions.

In Appendix~\ref{sec:symmetry_arguments}, assuming that the moving configurations satisfy the symmetries of the energy, it is shown that
\begin{equation}
    F_\mu := F_\mu^U = F_\mu^L,\quad F_2 = 0,\qquad\text{and}\qquad \Gamma_\mu := \Gamma_\mu^U = \Gamma_\mu^L,\quad \Gamma_1=0.
\end{equation}
Equations~\eqref{eq:dotR} become
\begin{equation} \label{eq:Papanicolaou}
\begin{split}
 \Skyrmion \dot{R}_\mu^U & = \alpha_1 \emn (F_\nu - \alpha \Gamma_\nu) \\
 \Skyrmion \dot{R}_\mu^L & = \alpha_1 \emn (F_\nu + \alpha \Gamma_\nu).
\end{split}
\end{equation}
These show that $R_x := R_x^U = -R_x^L,\; R_y := R_y^U = R_y^L$ and 
\begin{equation}
 \Skyrmion \dot{R}_1 = -\alpha_1\alpha \Gamma_2, \qquad
 \Skyrmion \dot{R}_2 = -\alpha_1 F_1.
\end{equation}
We can divide the two equations to get an equation for the trajectory,
\begin{equation}
    \frac{d\guidex}{d\guidey} =  \frac{\alpha\Gamma_2}{F_1}.
\end{equation}
It gives the angle $\theta$ of the trajectory of the skyrmion in either layer to the $y$-axis, 
\begin{equation} \label{eq:angle_of_motion_exact}
\tan\theta = \Big\lvert\frac{\alpha\Gamma_2}{F_1}\Big\rvert.
\end{equation}

In contrast to the approximate equation \eqref{eq:angle_of_motion}, the ratio on the right-hand side depends explicitly on $F_1$ and thus the angle of motion can be different on the outward and return journeys.
At the beginning of the motion, $F_1$ and $\Gamma_2$ can be calculated (see Appendix \ref{sec:initial_angle}) giving an initial angle of propagation that may be different from that in \eqref{eq:angle_of_motion}.

Meanwhile the argument in Sec. \ref{sec:general_Thiele_eqns} tells us that once the skyrmion velocities are changing slowly, provided the distortion is small, the angle is constant and equal to $\arctan(\alpha\Gamma/\Skyrmion)$ (ignoring signs). Comparing \eqref{eq:angle_of_motion} and \eqref{eq:angle_of_motion_exact}, we find that as this adiabatic approximation becomes more accurate, $\Gamma_2/F_1 \to \Gamma$. This in fact tells us that the ratio must change, given the initial value of $\Gamma_2/F_1$ is not equal to $\Gamma$ in general.

Gaining a better understanding of this initial transient regime would require modelling the time evolution of $\Gamma_2/F_1$ between these initial and limiting values, which is beyond the scope of this paper. However, we observe numerically that in the regime where $\Gamma_2/F_1$ notably changes, it oscillates around its `equilibrium' value of $\Gamma$. Thus even though there is some deviation from perfect remembering behaviour, it is small, as the times where $\Gamma_2/F_1>\Gamma$ are approximately cancelled by the times where $\Gamma_2/F_1<\Gamma$.

\subsection{Initial angle of propagation \label{sec:initial_angle}}

We can calculate $\Gamma_2$ at the moment when the magnetic field is switched on.
Since $\magnT$ is initially rotationally symmetric, we can simplify by going to polar co-ordinates $(r,\phi)$:
\begin{equation}
 \magnT = \begin{pmatrix}
    \sin\Theta(r)\cos(\phi+\gamma)\\
    \sin\Theta(r)\sin(\phi+\gamma)\\
    \cos\Theta(r)
\end{pmatrix}
\implies \Gamma_2 = \frac{1}{4\pi}\int_{r=0}^\infty \int_{\phi=0}^{2\pi}h(r\cos\phi)\sin^2\Theta(r)\cos\phi\, dr\, d\phi.
\end{equation}
We consider that $h(x_1)$ is an odd function, plus a constant, and thus $h'(x_1)$ is even. An overall constant in $h$ gives zero contribution in the integral for $\Gamma_2$.
In general, we can write
\begin{equation}
    \Gamma_2 = \frac{1}{2\pi}\int_{r=0}^\infty \int_{\phi=0}^{\pi}h^{-}(r\cos\phi)\sin^2\Theta(r)\cos\phi\, dr\, d\phi
\end{equation}
where $h^-$ is odd.
But more than that we need to use the fact that $h^-$ has basically positive gradient, being approximately linear in the area where it is most significant.

To go further, we specify the form of $h(x_1)$. For simplicity, we take 
\begin{equation}
    h(x_1)=\begin{cases}
        h_0 & x_1<-L\\
        h_0-g(x_1+L) & -L<x_1<L\\
        h_0-2g L & x_1>L
    \end{cases}.
\end{equation}
(We do not need to specify $h_0$ for our purposes but the simplest case would be $h_0=gL$.)
If $L$ is significantly larger than the skyrmion (in the simulations in the main text, the lengthscale of the gradient is between 5 and 10 times the skyrmion radius), then the integral can be well-approximated by
\begin{equation}
\Gamma_2 \simeq- \frac{g}{2} \int_0^\infty\sin^2\Theta(r)\, r\,dr .
\end{equation}
We can then calculate the predicted initial angle, which is related to $\Gamma_2/F_1$ through \eqref{eq:angle_of_motion_exact}. By a similar calculation,
\begin{equation}
    F_1 \simeq g\int_0^\infty(1-\cos\Theta)rdr 
\end{equation}
again assuming $L$ is greater than the skyrmion radius. The force $F_1$ is then the `Zeeman susceptibility' of the skyrmion, the derivative of the Zeeman energy with respect to the coupling constant. Then
\begin{equation}
   \tan\theta|_{t=0} = \frac{\alpha\lvert\Gamma_2\rvert}{\lvert F_1\rvert}\Big|_{t=0} = \frac{\alpha\int_0^L \sin^2\Theta\, r dr}{\int_0^L (1-\cos\Theta)rdr};
\end{equation}
the ratio of $\Gamma_2$ to $F_1$ at the initial time (and thus the initial angle of propagation) is a function of the ratio of Zeeman and anisotropy energy susceptibilities. By contrast, $\Gamma$ at the initial time is a function of the exchange energy. There is no reason for these to be the same.
Indeed, we can see they are not the same in the limit of a skyrmion with large radius $R\gg1$ (while still smaller than the lengthscale of field gradient $L$): we assume that the skyrmion domain wall is $\cos\Theta = \tanh(r-R)$ and have (set $T=r-R$)
\begin{align*}
& \int_0^L \sin^2\Theta\,r dr \approx R \int_{-\infty}^\infty \sech^2T\,dT = 2R \\
& \int_0^L (1-\cos\Theta)\,r dr \approx \int_{-R}^0 (1-\tanh T) (R+T)\,dT \approx  R^2.
\end{align*}
We have
\[
\frac{\Gamma_2}{F_1} = -\frac{2}{R} \implies \theta \to\frac{2\alpha}{ R} .
\]
To compare with the quasi-steady-state angle for skyrmion propagation, we should also calculate $\Gamma$. We find that
\begin{align*}
\Gamma \sim 2\pi R \implies \theta\to \arctan(2\pi\alpha R)\to\frac{\pi}{2}
\end{align*}
The final angle becomes close to $\frac{\pi}{2}$ as $R\to\infty$, while the initial angle becomes close to $0$.

\bigskip
\bibliographystyle{apsrev4-2.bst}
\bibliography{references.bib}

\end{document}

%% file: preamble.tex
\usepackage[margin=1in]{geometry} 
\usepackage{amsmath,amssymb}
\usepackage{amsfonts}
\usepackage{mathrsfs}
\usepackage{hyperref}
\usepackage{enumerate}
\usepackage{verbatim}
\usepackage{color}
\usepackage{bm}
\usepackage{cancel}
\usepackage{ulem}
\definecolor{green}{RGB}{0,255,0}
\definecolor{red}{RGB}{255,0,0}
\usepackage{enumitem}
\setlist{topsep=7pt,itemsep=-2pt}
\usepackage{graphicx}

%% file: symbols.tex
\newcommand{\Magn}{\mathbf{M}}
\newcommand{\magn}{\mathbf{m}}

\newcommand{\magnT}{\mathbf{U}}
\newcommand{\magnB}{\mathbf{L}}

\DeclareMathOperator{\sech}{sech}
\newcommand{\p}{\partial}
\newcommand{\emn}{\epsilon_{\mu\nu}}

\newcommand{\e}{\bm{\hat{e}}}

\newcommand{\ldw}{\ell_{\rm w}}
\newcommand{\Energy}{E}
\newcommand{\Etotal}{E_{\rm tot}}

\newcommand{\Eint}{E_{\rm int}}
\newcommand{\tmagn}{\mathcal{M}}

\newcommand{\heff}{\bm{f}}
\newcommand{\hext}{h}
\newcommand{\bhext}{\bm{h}}

\newcommand{\bHext}{\bm{H}}

\newcommand{\DM}{D}
\newcommand{\dm}{\lambda}

\newcommand{\interJ}{j_c}
\newcommand{\Dcoupling}{A_c}
\newcommand{\Anisotropy}{K}

\newcommand{\skyrmion}{q}
\newcommand{\Skyrmion}{\mathcal{Q}}

\newcommand{\guidex}{R_1}
\newcommand{\guidey}{R_2}
\newcommand{\vel}{\upsilon}